\documentclass[9pt,twocolumn,twoside]{osajnl}

\journal{ol} 

\setboolean{shortarticle}{true}

\title{Low coherence-induced resonance in double-layer structures having parity-time symmetry}

\author[1*]{P. A. Brand\~ao}
\author{J. P. Mendon{\c{c}}a}
\author{S. B. Cavalcanti}

\affil{Instituto de F\'isica, Universidade Federal de Alagoas, Macei\'o, 57072-900, Brazil.}

\affil[*]{Corresponding author: paulo.brandao@fis.ufal.br}

\begin{abstract}
We derive simple formulae for the transmittance $T$ and reflectance $R$ of Gaussian-Schell beams incident upon any stratified dielectric structure by using second-order classical coherence theory in the space-frequency picture. The formalism is applied to a particular structure consisting of a double-layer, with balanced gain and loss, satisfying the parity-time symmetry conditions. It is shown that sources with a low degree of spatial coherence, on the order of the wavelength, can induce large resonant peaks in the transmitted and reflected amplitudes. The resonance peaks vanish as the spatial coherence increases.
\end{abstract}

\setboolean{displaycopyright}{true}

\begin{document}

\maketitle

Thirty years ago, W. Wang and E. Wolf addressed the problem of transmission and reflection properties of partially coherent beams through a general stratified medium \cite{wang1990effects}. The examples given by the authors relied on the space-frequency formalism of classical coherence theory and on the quasi-homogeneous approximation for the source field. Also, their definitions for ``reflectance'' and ``transmittance'' relied on the far-zone approximation for the transmitted and reflected fields. Experimental results confirming some aspects of the theory were published shortly after the formulation of the theoretical approach \cite{vaishya1993effect}. Unfortunately, their work did not received much attention through the years despite the interesting results they obtained, such as the strong dependence of the reflection and transmission 
 coefficients on the spatial coherence properties of the incident field. Spectral changes in the transmitted and reflected beams were also considered later \cite{wang1992changes,chen1996far}. 

Since the seminal work of C. M. Bender and S. Boettcher on non-Hermitian Hamiltonians having parity-time symmetry \cite{bender1998real}, optics has been passing through a profound transformation regarding the fundamental properties of the interaction between radiation and matter \cite{feng2017non,longhi2018parity}. This was initially achieved due to an analogy existent between the paraxial wave equation of optics and the time-dependent Schr\"odinger equation. The complex, quantum, non-Hermitian Hamiltonian is mapped into a complex, classical, refractive index whose imaginary part is related to gain and loss properties of the material medium. Despite the large amount of new effects discovered in non-Hermitian optical settings, the role of classical optical coherence, and how it influences the interaction between radiation and non-Hermitian matter, has just begun to be explored in scattering systems \cite{brandao2019non,brandao2019scattering,pinto2020asymmetrical,Vieira_2020,brandao2020scattering}. It was found that deterministic non-Hermitian materials can drastically alter the directions \cite{brandao2019scattering} and the frequency spectrum of the scattered radiation, giving rise to a non-Hermitian Wolf effect \cite{Vieira_2020}. A recent generalization of light scattering from random, parity-time-symmetric materials, has demonstrated that a new class of localized, statistically stationary materials, can be constructed from parity-time symmetry concepts \cite{brandao2020scattering}. It now seems evident that the combination of classical coherence theory with non-Hermitian concepts give rise to many intriguing phenomena yet unexplored. With this in mind, the objective of this paper is to generalize the problem of transmission and reflection properties of (spatially) partially coherent beams through dielectric layers containing gain and loss, satisfying the conditions of parity-time symmetry. Furthermore, we intend to observe how the spatial coherence properties of the incident beam influences the transmittance and reflectance of the structure.

Let us begin by considering a homogeneous, isotropic, linear and planar material, described by a piecewise constant complex refractive index $n(z) = n_R(z) + in_I(z)$, positioned in the interval $ 0 \leq z \leq D$, embedded in vacuum. A stochastic incident scalar monochromatic light beam $\psi_i(x,z,\omega)$, arriving from vacuum ($z<0$), strikes the material at $z = 0$. We assume that all field quantities are independent of the $y$ direction. As a consequence, two beams are originated outside the material: a reflected  beam, $\psi_r(x,z,\omega)$, and a transmitted one, $\psi_t(x,z,\omega)$. We write each beam as a sum of plane waves (angular spectrum) and omit the frequency dependence on $\omega$ from now on,
\begin{equation}
\begin{split}
    \psi_{i}(x,z) &= \int_{-\infty}^{\infty}\psi_{i}(k_x)e^{ik_xx}e^{iz\sqrt{k^2-k_x^2}}dk_x, \quad (z \leq 0), \\
    \psi_{r}(x,z) &= \int_{-\infty}^{\infty}\psi_{r}(k_x)e^{ik_xx}e^{-iz\sqrt{k^2-k_x^2}}dk_x, \quad (z \leq 0), \\
    \psi_{t}(x,z) &= \int_{-\infty}^{\infty}\psi_{t}(k_x)e^{ik_xx}e^{iz\sqrt{k^2-k_x^2}}dk_x, \quad (z \geq D),
\end{split}
\end{equation}
where $k = \omega/c$ is the wavenumber with $c$ being the speed of light in vacuum and $\psi_l(k_x)$ is the Fourier transform of $\psi_l(x,0)$:
\begin{equation}
    \psi_l(k_x) = \frac{1}{2\pi}\int_{-\infty}^{\infty} \psi_l(x,0)e^{-ik_x x} dx \quad (l = i,r,t).
\end{equation}
Here, for the sake of simplicity, the incident field $\psi_i$ is a beam normally incident on the interface. It is possible to generalize the analysis by considering a spatial displacement of all plane waves that compose the incident angular spectrum so that the incidence becomes oblique  \cite{son2015light}. 
The continuity of the fields and their normal derivatives imposed at the interfaces $z = 0$ and $z = D$ connects all spectral amplitudes. Here, our interest is in the transmitted spectral amplitude $t(k_x)$ at wavevector $k_x$, defined by

\begin{equation}\label{tspec}
        t(k_x) = \frac{\psi_t(k_x)}{\psi_i(k_x)}.
\end{equation}
The transmitted beam can then be written as
\begin{equation}\label{psit}
    \begin{split}
        \psi_{t}(x,z) &= \int_{-\infty}^{\infty}\psi_{t}(k_x)e^{ik_x x}e^{iz\sqrt{k^2-k_x^2}}dk_x, \\
        &= \int_{-\infty}^{\infty}t(k_x)\psi_{i}(k_x)e^{ik_x x}e^{iz\sqrt{k^2-k_x^2}}dk_x,
    \end{split}
\end{equation}
and its statistical properties can be calculated from the cross-spectral density
\begin{equation}\label{Wt}
    W_t(x_1,x_2,z_0) = \langle \psi_t^*(x_1,z_0)\psi_t(x_2,z_0) \rangle_{\omega},
\end{equation}
obtained from averaging \eqref{psit} over monochromatic realizations at a fixed plane $z = z_0$. From the cross-spectral density, one can obtain the spectral density $S_t(x,z) = W_t(x,x,z)$ of the transmitted radiation and we define the \textit{transmittance} $T(z_0)$ at the plane $z = z_0$ as
\begin{equation}\label{trans}
    T(z_0) = \sqrt{2\pi}\frac{\int_{-\infty}^{\infty} S_t(x,z_0) dx}{\int_{-\infty}^{\infty} S_i(x,0) dx},
\end{equation}
where $S_i(x,0) = W_i(x,x,0)$ is the spectral density of the incident field at $z = 0$ and the $\sqrt{2\pi}$ factor is included to guarantee that $T = 1$ when $|t(k_x)| = 1$ as we shall explain shortly after the transmittance is defined below. 

To relate the spectral density $S_t(x,z)$ of the transmitted beam with the statistical properties of the incident field at $z = 0$, described by the cross-spectral density $W_i(x_1,x_2,0)$, we substitute \eqref{psit} into \eqref{Wt}:
\begin{equation}\label{wt11}
    \begin{split}
        &W_t(x_1,x_2,z_0) = \iint_{-\infty}^{\infty}dk_{x1}dk_{x2} W_i(k_{x1},k_{x2}) t^*(k_{x1}) t(k_{x2}) \\
        &\times \exp{[i(x_2k_{x2}-x_1k_{x1})]} \exp{\left[ iz_0 \left(\sqrt{k^2-k_{x2}^2} - \sqrt{k^2-k_{x1}^2}\right)\right]},
    \end{split}
\end{equation}
where the \textit{spectral} correlation function $W_i(k_{x1},k_{x2})$ that appears in  \eqref{wt11} is related to $W_i(x_1,x_2,0)$ by the formula
\begin{equation}\label{Wi}
    \begin{split}
    W_i(k_{x1},k_{x2}) &= \langle \psi_i^*(k_{x1})\psi_i(k_{x2}) \rangle_{\omega} \\
    &= \frac{1}{(2\pi)^2} \iint dx_{1} dx_2 W_i(x_1,x_2,0)\\
    &\times\exp[-i(k_{x2}x_2 - k_{x1}x_1)],
    \end{split}
\end{equation}
where we have used the Fourier inverse theorem. The spectral density $S_t(x,z_0)$ of the transmitted beam is finally obtained from \eqref{wt11} by setting $x_1 = x_2 = x$:
\begin{equation}\label{St2}
    \begin{split}
        &S_t(x,z_0) =\iint_{-\infty}^{\infty}dk_{x1}dk_{x2} W_i(k_{x1},k_{x2}) t^*(k_{x1}) t(k_{x2}) \\
        &\times \exp{[ix(k_{x2}-k_{x1})]} \exp{\left[ iz_0 \left(\sqrt{k^2-k_{x2}^2} - \sqrt{k^2-k_{x1}^2}\right)\right]}.
    \end{split}
\end{equation}
Notice that the numerator of \eqref{trans} can be written in the form
\begin{equation}\label{stsimple}
    \int_{-\infty}^{\infty}S_t(x,z_0)dx = 2\pi \int_{-\infty}^{\infty} S_i(k_x) |t(k_x)|^2 dk_x,
\end{equation}
which is independent of $z_0$. To obtain \eqref{stsimple}, \eqref{St2} was used and $S_i(k_x) = W_i(k_x,k_x)$.

We assume that the coherence properties of the incident beam are well described by a Gaussian-Schell model \cite{Mandel-Wolf},
\begin{equation}\label{Wi2}
    W_i(x_1,x_2,0) = \sqrt{S_i(x_1,0)}\sqrt{S_i(x_2,0)}\mu_i(x_1-x_2,0) ,
\end{equation}
with
\begin{equation}\label{Smu}
\begin{split}
    S_i(x,0) &= S_0\exp\left( \frac{-x^2}{\delta^2} \right), \\
    \mu_i(x_1-x_2,0) &= \exp\left[ \frac{-(x_1-x_2)^2}{2\Delta^2} \right].
\end{split}
\end{equation}
The function $\mu_i(x_1,x_2,0)$ is the spectral degree of coherence of the incident beam, $S_0$ is a constant amplitude, $\delta$ is related to the beam's width at $z = 0$ and $\Delta$ is the coherence length at $z = 0$. The usual monochromatic, spatially coherent, Gaussian beam is obtained directly from the above formulas in the limit $\Delta \rightarrow \infty$, implying that all points located in the transverse plane are fully correlated. Direct substitution of \eqref{Smu} into \eqref{Wi2} and then into \eqref{Wi} yields 
\begin{equation}\label{Wik}
\begin{split}
    W_i(k_{x1},k_{x2}) &= \frac{S_0 \delta^ 2\Delta}{(2\pi)^{3/2} \sqrt{\Delta^ 2 + 2\delta^2}}\exp\left[\frac{ -\frac{\Delta^2}{2}(k_{x1}^2 + k_{x2}^2)}{2 + \frac{\Delta^2}{\delta^2}} \right] \\
    &\times\exp\left[ \frac{- \frac{\delta^2}{2}(k_{x1} - k_{x2})^ 2}{2 + \frac{\Delta^2}{\delta^2}} \right].
\end{split}
\end{equation}
In the limit $\Delta \rightarrow \infty$,  \eqref{Wik} becomes separable, as expected. The function $S_i(k_x)$ is readily obtained from \eqref{Wik}:
\begin{equation}\label{Sik}
    S_i(k_{x}) = \frac{S_0 \delta^ 2\Delta}{(2\pi)^{3/2} \sqrt{\Delta^ 2 + 2\delta^2}} \exp\left(\frac{ -\Delta^2k_{x}^2}{2 + \frac{\Delta^2}{\delta^2}} \right).
\end{equation}
Then, \eqref{trans} can be rewritten by using \eqref{stsimple} and \eqref{Sik} as
\begin{equation}\label{transfinal}
\begin{split}
T &= \frac{2\pi\sqrt{2}}{S_0\delta}\int_{-\infty}^{\infty} S_i(k_x)|t(k_x)|^2dk_x \\
     &=\frac{\delta\Delta}{\sqrt{\pi}\sqrt{\Delta^2 + 2\delta^2}}\int_{-\infty}^{\infty}\exp\left(\frac{ -\Delta^2k_{x}^2}{2 + \frac{\Delta^2}{\delta^2}} \right) |t(k_x)|^2 dk_x.
\end{split}
\end{equation}
The transmittance $T$ expressed by \eqref{transfinal} is the main result of our analysis. It is important to emphasize that it is generally valid, independent of the type of dielectric interfaces existent in the interval $[0,D]$. The only restriction imposed on \eqref{transfinal} is that the incident field be a Gaussian-Schell beam described by \eqref{Smu}. The reader can easily verify that for $|t(k_x)|^2 = 1$, \eqref{transfinal} results in $T = 1$, which justifies the inclusion of the factor $\sqrt{2\pi}$ in the definition of $T$ [\eqref{trans}]. Thus, once the spectral amplitude $t(k_x)$ for a specific layered structure is known, a direct integration can be performed to obtain the transmittance properties of the dielectric. 

The \textit{reflectance} $R$ can be defined in the same way as the transmittance through the definition of the reflected spectral amplitude
\begin{equation}
    r(k_x) = \frac{\psi_r(k_x)}{\psi_i(k_x)}.
\end{equation}
By performing the same analysis as before, we obtain the formula
\begin{equation}\label{reffinal}
R = \frac{\delta\Delta}{\sqrt{\pi}\sqrt{\Delta^2 + 2\delta^2}}\int_{-\infty}^{\infty}\exp\left(\frac{ -\Delta^2k_{x}^2}{2 + \frac{\Delta^2}{\delta^2}} \right) |r(k_x)|^2 dk_x.
\end{equation}
For conservative (Hermitian) systems where $|t(k_x)|^2 + |r(k_x)|^2 = 1$ is satisfied for every transverse wavevector $k_x$, it is readily found that $T+R = 1$, as expected.

We are now in a position to consider specific results of numerical simulations involving the transmission of a Gaussian-Schell beam through non-Hermitian structures. Our main task is to understand how the transmittance [\eqref{transfinal}], and the reflectance [\eqref{reffinal}], depend on the non-Hermitian parameter $n_I$ as well as on the coherence length $\Delta$ of the Gaussian-Schell beam. To this end, it only remains to obtain the transmitted and reflected spectral amplitudes $t(k_x)$ and $r(k_x)$. This is easily achieved by using the transfer matrix $\pmb{M}$ approach for the system under consideration \cite{markos2008wave}. Suppose we find that the $\pmb{M}$-matrix has the form
\begin{equation}
    \pmb{M}(k_x) = \begin{bmatrix} M_{11}(k_x) & M_{12}(k_x) \\
    M_{21}(k_x) & M_{22}(k_x) \end{bmatrix},
\end{equation}
with $M_{ij}$ being complex-valued entries. Then, the spectral amplitudes are given by $t(k_x) = \text{det}[\pmb{M}(k_x)]/M_{22}(k_{x})$ and  $r(k_x) = -M_{21}(k_x)/M_{22}(k_x)$ where $\text{det}[\pmb{M}(k_x)] = M_{11}(k_x)M_{22}(k_x) - M_{12}(k_x)M_{21}(k_x)$ is the determinant of the transfer matrix \cite{markos2008wave}.

In order to apply the above formalism, let us consider a double-layer system having parity-time symmetry. By this we mean a dielectric layer with refractive index $n_1 = n_R + in_I$, with $n_R$ and $n_I$ real-valued, occupying the region $[0,L]$ and another layer with refractive index $n_2 = n_1^*$ occupying the region $[L,2L]$. The total length of the system being $D = 2L$ and we assume both layers to be embedded in vacuum ($n = 1$). These parity-time-symmetric double-layer structures have attracted a great deal of attention due to its unusual response to optical fields. Effects such as unconventional lasing modes \cite{ge2011unconventional}, invisibility \cite{mostafazadeh2013invisibility,shen2014unidirectional}, complete transmission through \textit{epsilon-near-zero} layers \cite{savoia2014tunneling} and nonlinear scattering \cite{mostafazadeh2019nonlinear} have been considered recently. 
The elements $M_{ij}$ of the transfer matrix $\pmb{M}$ for periodic media with arbitrary number of layers were derived in \cite{hoenders2005coherence}.

\begin{figure}[htbp]
\centering
\includegraphics[width=0.9\linewidth]{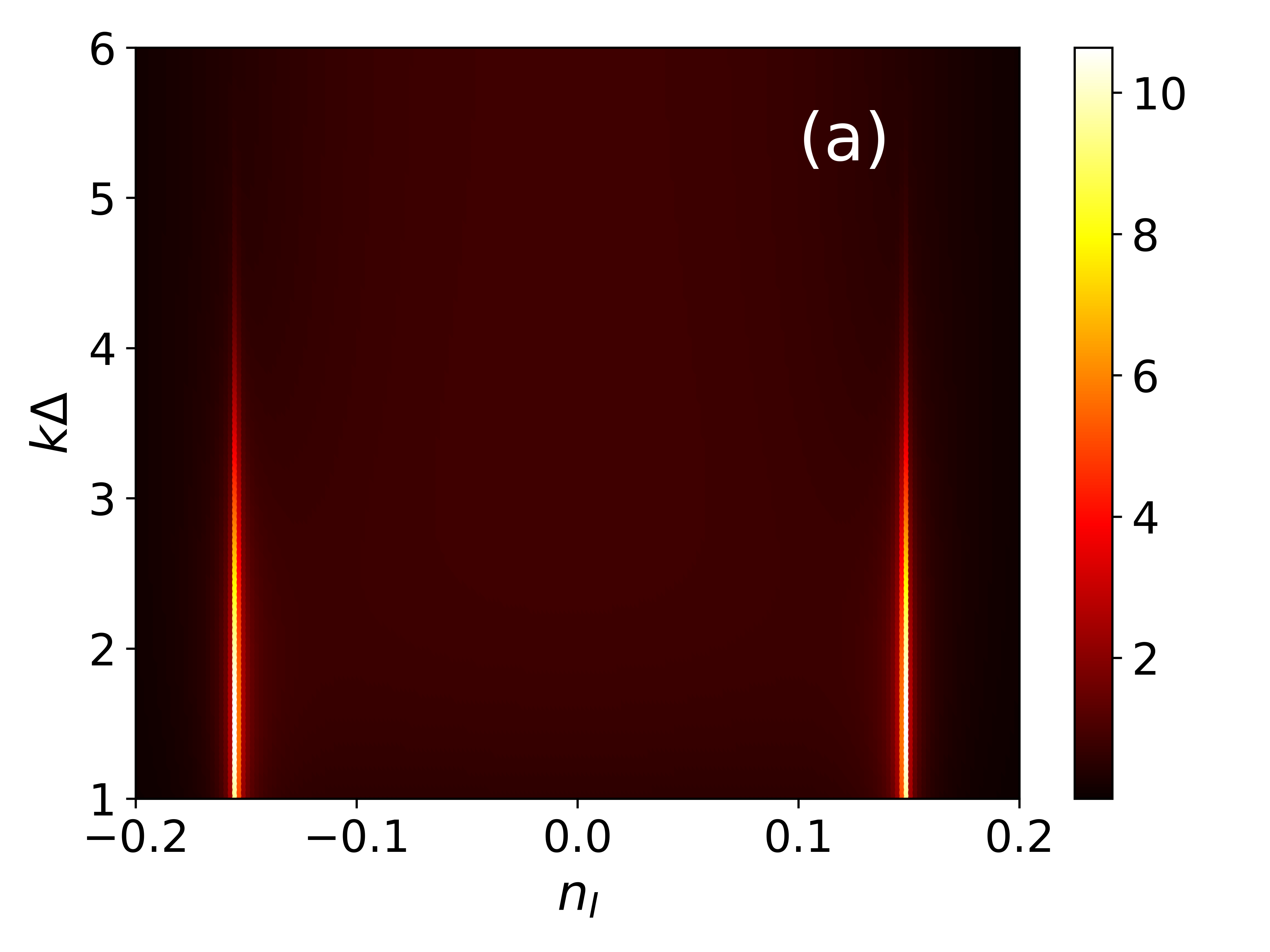}
\includegraphics[width=0.9\linewidth]{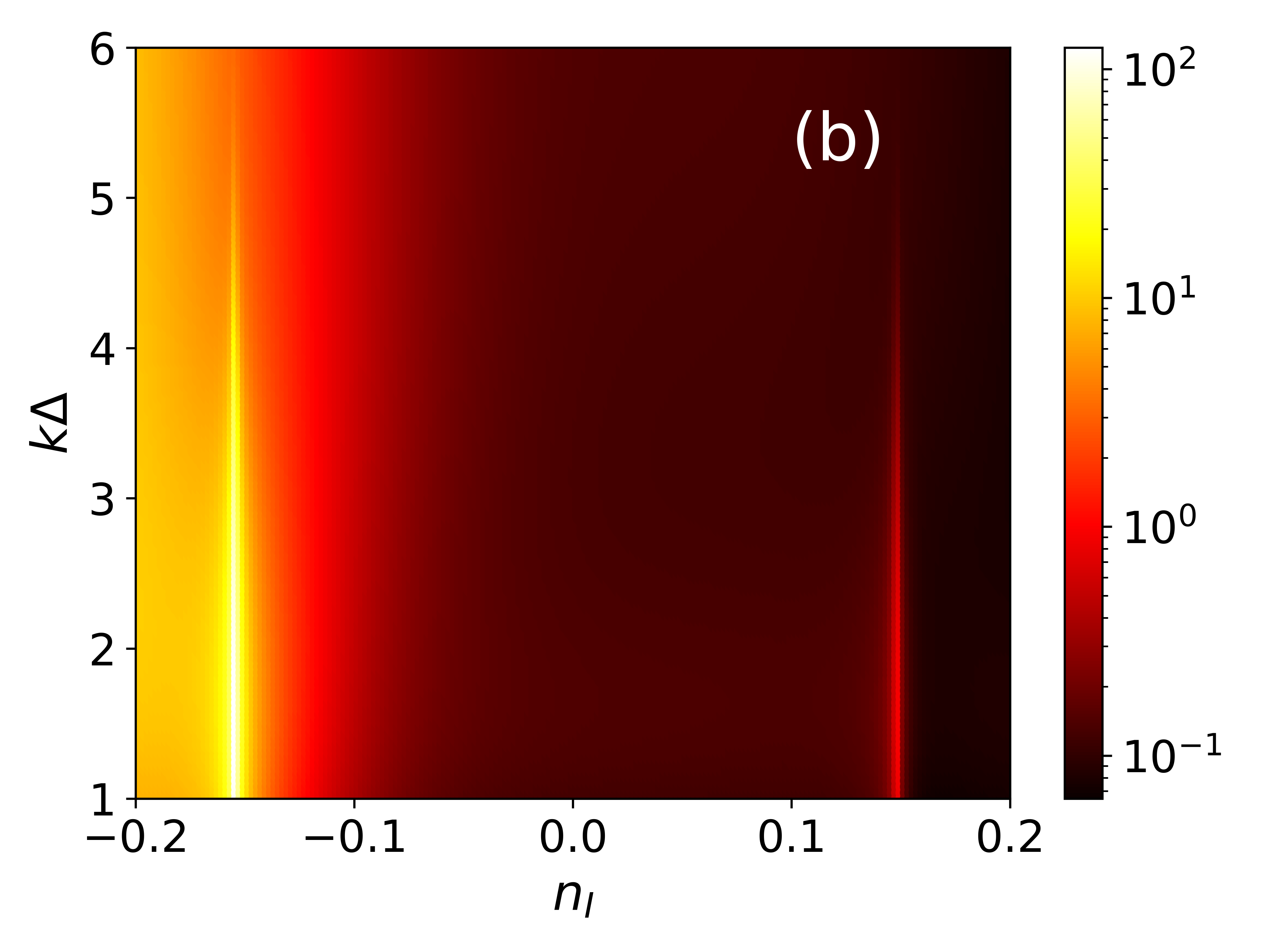}
\caption{(a) Transmittance $T$ and (b) reflectance $R$ obtained from \eqref{transfinal} and \eqref{reffinal}, respectively, as functions of the coherence length $k\Delta$ of the incident beam for various values of $n_I$. Parameters used: $k\delta = 50$}, $kL = 10$ and $n_R = 1.5$. In the reflectance density plot (b) a logarithm scale is used.
\label{fig1}
\end{figure}

To start, let us choose $n_R = 1.5$, $k_x \rightarrow k_x/k$, $kL = 10$ and $k\delta = 50$ and investigate how $R$ and $T$ depend on the parameters $\Delta$ and $n_I$ since they are the most relevant ones to our analysis. Figure \ref{fig1}(a) shows a density plot of $T(n_I,k\Delta)$ for $n_I \in [-0.2,0.2]$ and $k\Delta \in [1,6]$. It is readily seen from this plot that there are two strong resonant transmission peaks (with $T \sim$ 10) located around $n_I \approx \pm 0.151$ that vanish as $k\Delta$ increases (large spatial coherence). One can verify from these plots that, when we fix the width of the incident beam, the transmitted peaks are only evident in the low coherence regime of the source, more specifically, for sources whose coherence length is of the order of the wavelength. The transmittance is symmetric with respect to the inversion round $n_I = 0$. 

\begin{figure}[htbp]
\centering
\includegraphics[width=\linewidth]{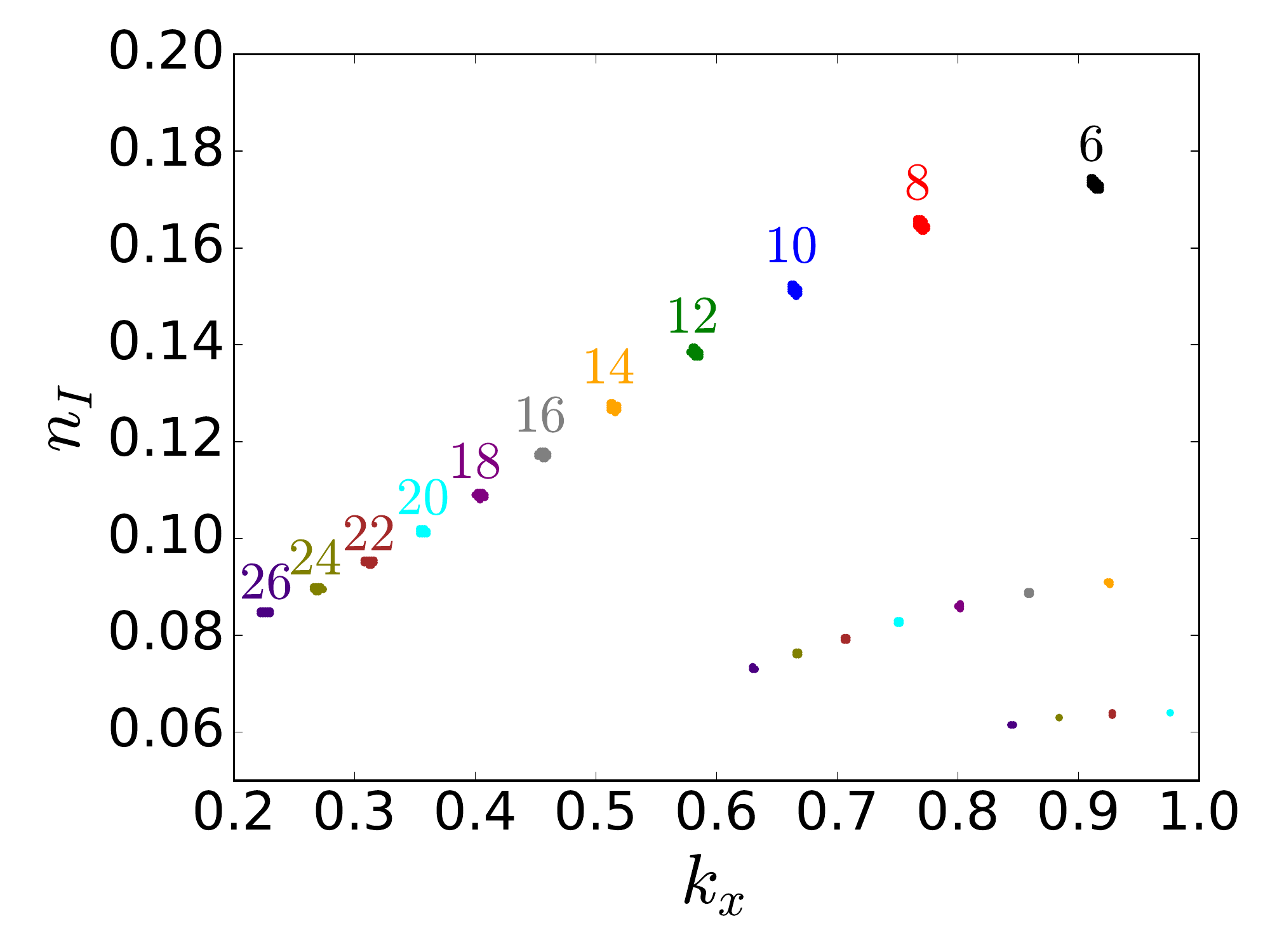}
\caption{(color online) Position of the peaks ($|t(k_x)|^2>500$) in the positive $k_x$--$n_I$ plane for various values of $L$ and $n_R=1.5$. Note that the number of peaks increases as $L$ increases. Both $|t|^2$ and $|r|^2$ exhibit a symmetric behavior under the operation $k_x\rightarrow -k_x$.}
\label{fig2}
\end{figure}

\begin{figure}[htbp]
\centering
\includegraphics[width=\linewidth]{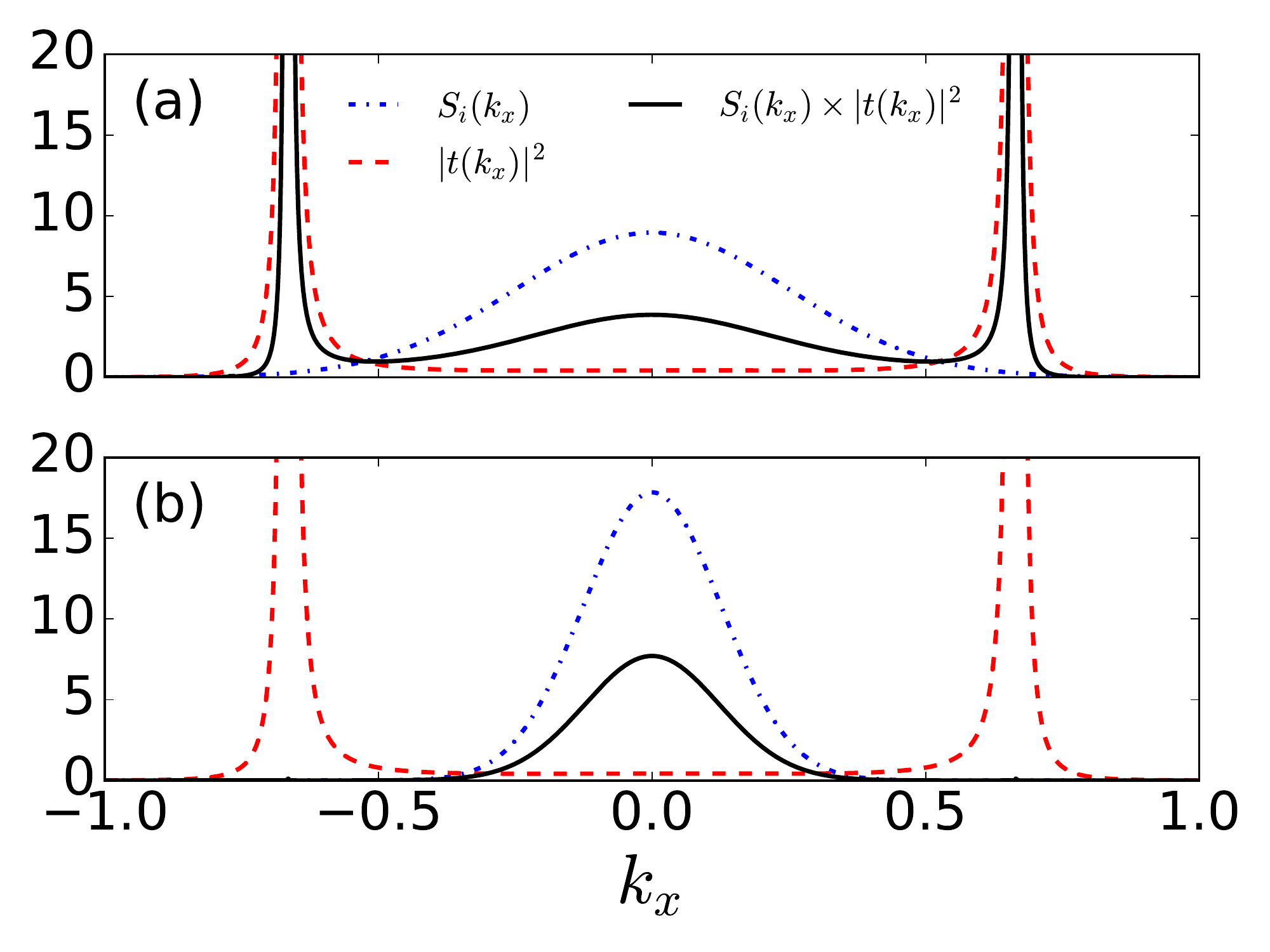}
\caption{Functions of the medium $|t(k_x)|^2$, of the incident beam $S_i(k_x)$, and their product. (a) $k\Delta=4$, (b) $k\Delta=8$. Note that the coherence length controls the width of $S_i$ and consequently, the overlap between these two functions which is responsible for the resulting strong resonance. Parameters used: $S_0=1$, $k\delta = 50$, $kL = 10$, $n_R = 1.5$, and $n_I=0.151$} 
\label{fig3}
\end{figure}

The reflectance displays a more asymmetric character, as can be visualized in part (b) of Fig. \ref{fig1}. The scale is logarithm for better visualization since the peak located at $n_I \approx - 0.151$ is much higher than the one located at $0.151$. This asymmetric effect in the reflectance is characteristic of non-Hermitian systems. Notice that exchanging the imaginary parts of both, $n_1$ and $n_2$, is equivalent to the assumption that the incident field is striking the first interface from the other side of the structure. Therefore, the formalism clearly contains non-Hermitian features. 


To gain a deeper understanding of the results shown in Fig. \ref{fig1} and to elucidate the origin of these resonances, we now proceed to the examination of the interaction between the incident ensemble of realizations of plane waves, whose properties are represented by the spectral function $S_i(k_x)$, and the spectral amplitudes $|t(k_x)|^2$ and $|r(k_x)|^2$. Let us consider the spectral amplitudes first. Figure \ref{fig2} shows the plot of the peaks of $|t(k_x)|^2$ as a function of $k_x$ and positive values of $n_I$ for several values of $kL$. The dots correspond to the respective values of the labels along with the colours. For example, for $kL = 6$ (black dot) we see that there is only one resonant peak in this interval, located around $(k_x,n_I) = (0.92,0.17)$. As $kL$ increases, more resonant peaks appear. For $kL = 20$, for example, the cyan dots indicate that there are three strong peaks in the $(k_x,n_I)$ plane and so on. Clearly, the behavior of the peaks as a function of $n_I$ and $kL$ is nontrivial and a closed form expression for their positions is very difficult to derive. We performed a numerical analysis to obtain the peak locations shown in Fig. \ref{fig2}. Although only positive values of $k_x$ and $n_I$ is considered in this plot, we verified that the entire peak structure of $|t|^2$ can be obtained by a simultaneous reflection of Fig. \ref{fig2} about $k_x = 0$ and $n_I = 0$. The peak locations of the reflected spectral amplitude $|r|^2$ are the same as the locations of the transmitted ones except for, they only occur for negative values of $n_I$.

Now that we have verified the locations of the resonant peaks of the structure, it is possible to understand the reason behind the strong resonances in the transmittance $T$ and reflectance $R$ and why they may disappear depending on the coherence properties of the incident beam. A glimpse at Eqs. (\ref{transfinal}) and (\ref{reffinal}) reveals that the important quantity is the product between $S_i(k_x)$, given by \eqref{Sik}, and the spectral amplitudes. The important parameter to consider in the integration is the width $\Gamma$ of $S_i(k_x)$, given by $\Gamma = \sqrt{\frac{2}{\Delta^2} + \frac{1}{\delta^2}}.$ As an example, suppose that we fix $\delta = 50$ and consider the behavior of $S_i(k_x)$, $|t(k_x)|^2$ and their product separately, as functions of $k_x$ for fixed $n_I = 0.151$. Part (a) of Figure \ref{fig3} depicts the plots of the three terms with $k\Delta = 4$.  In this case, since the width $\Gamma$ of the Gaussian $S_i(k_x)$ overlaps the resonances at $k_x \approx 0.6$, the product, represented by the continuous black line, exhibits very pronounced peaks. On the other hand, by increasing the spatial coherence to $k\Delta = 8$, part (b) of Fig. \ref{fig3} shows that the Gaussian width $\Gamma$ has almost no overlap with the resonances of $|t|^2$ so that the product between $S_i(k_x)$ and $|t|^2$ has no pronounced resonances. The non-existence of the overlap explains why the resonant peaks disappear when the spatial coherence of the incident field increases. The same considerations apply to the reflectance $R$, of course. We should also point out that, due to the form of the $\Gamma$ function, strong transmitted and reflected resonances are also expected when the incident beam is characterized by a high degree of spatial coherence. However, in this case, to observe the resonances, the width of the input spectral density must be small. Therefore, the ratio of the transverse coherent length to the beam width, i.e., $\frac{\Delta}{\delta}$ plays an important role in the enhancement of resonances.

In closing, we remark that the formalism developed here can be easily generalized to (2+1) dimensions, meaning that one can write for the incident, reflected and transmitted beams expressions such as $\psi_j(\pmb{\rho},z)$ with $\pmb{\rho} = (x,y)$ being a transverse vector in the plane. This generalization allows one to consider more exotic incident beams, possessing nontrivial phases and amplitudes, and its interaction with non-Hermitian structures could be studied from the point of view of classical coherence theory. All these final considerations are currently under investigation and will be published elsewhere.

\section*{Acknowledgments}
J. P. Mendonça thanks CAPES (Coordenação de Aperfeiçoamento de Pessoal de Nível Superior) and S. B. Cavalcanti thanks CNPq (Conselho Nacional de Desenvolvimento Científico e Tecnológico) for financial support.

\section*{Disclosures}

The authors declare no conflicts of interest.

\bibliography{sample}

\bibliographyfullrefs{sample}


\ifthenelse{\equal{\journalref}{aop}}{%
\section*{Author Biographies}
\begingroup
\setlength\intextsep{0pt}
\begin{minipage}[t][6.3cm][t]{1.0\textwidth} 
  \begin{wrapfigure}{L}{0.25\textwidth}
    \includegraphics[width=0.25\textwidth]{john_smith.eps}
  \end{wrapfigure}
  \noindent
  {\bfseries John Smith} received his BSc (Mathematics) in 2000 from The University of Maryland. His research interests include lasers and optics.
\end{minipage}
\begin{minipage}{1.0\textwidth}
  \begin{wrapfigure}{L}{0.25\textwidth}
    \includegraphics[width=0.25\textwidth]{alice_smith.eps}
  \end{wrapfigure}
  \noindent
  {\bfseries Alice Smith} also received her BSc (Mathematics) in 2000 from The University of Maryland. Her research interests also include lasers and optics.
\end{minipage}
\endgroup
}{}

\end{document}